# Analytic prognostic for petrochemical pipelines

Abdo Abou Jaoude[1], Seifedine Kadry[2], Khaled El-Tawil[3],
Hassan Noura[4], and Mustapha Ouladsine[5]

[1]Ph.D. student at Paul Cézanne University, Aix-Marseille, France, and the Lebanese University (EDST), Lebanon, abdoaj@idm.net.lb
[2]Associate professor, American University of the Middle East, Kuwait, skadry@gmail.com
[3]Associate Professor, Lebanese University, Faculty of Engineering & EDST, Campus of Hadath, Lebanon, Khaled_tawil@ul.edu.lb
[4]Professor, United Arab Emirates University, Department of Electrical Engineering, hnoura@uaeu.ac.ae
[5]Professor, Laboratoire des Sciences de l'Information et des Systèmes (LSIS, UMR CNRS 6168),
Paul Cézanne University, Aix-Marseille, France, mustapha.ouladsine@lsis.org

**Pipelines tubes are part of vital mechanical systems largely used in petrochemical industries. They serve to transport natural gases or liquids. They are cylindrical tubes and are submitted to the risks of corrosion due to high PH concentrations of the transported liquids in addition to fatigue cracks due to the alternation of pressure-depression of gas along the time, initiating therefore in the tubes' body micro-cracks that can propagate abruptly to lead to failure. The development of the prognostic process for such systems increases largely their performance and their availability, as well decreases the global cost of their missions. Therefore, this paper deals with a new prognostic approach to improve the performance of these pipelines. Only the first mode of crack ($K=K_I$), i.e. the opening mode, is considered.**

**Key words:** Analytic laws, trajectory of degradation, pipelines, fatigue cracks, prognostic, remaining lifetime.

## INTRODUCTION

PROGNOSTIC DEFINITION

The term prognostic founds its origin in the Greek word "progignôskein" which means "to know in advance". Industrial Prognostic is called the prediction of a system's lifetime and corresponds to the last level of the classification of damage detection methods introduced by [1]. Prognostic can also be defined as a probability measure: a way to quantify the chance that a machine operates without a fault or failure up to some future time. This "probabilistic prognostic value" is all the more an interesting indication as the fault or failure can have catastrophic consequences (e.g. nuclear power plant) and maintenance manager need to know if inspection intervals are appropriate. However, a small number of papers address this acceptation for prognostic [2], [3].

Finally, although there are some divergences in literature, prognostic can be defined as: "prognostic is the estimation of time to failure and risk for one or more existing and future failure modes" [4]. In this acceptation, prognostic is also called the "prediction of a system's lifetime" as it is a process whose objective is to predict the remaining useful life (RUL) before a failure occurs given the current machine condition and past operation profile [5]. The main steps defined in this standard are summarized in Figure 1.

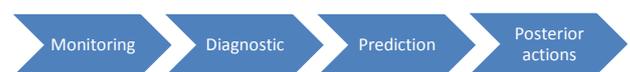

**Figure 1.** Summary of the ISO 13381-1: 2004 standard main steps

The first step consists in monitoring the system by a set of sensors or inspections achieved by operators. The monitored data are then pre-processed in order to be used by the Diagnostic module. The output of this module identifies the actual operating mode. This state is then projected in the future, by using adequate tools, in order to predict the system's future state. The intersection point between the value of each projected parameter or feature and its corresponding alarm threshold leads to what is known as RUL (Remaining Useful Life) of the system (Figure 2). Finally, appropriate maintenance actions can be taken depending on the estimated RUL. These actions may aim at eliminating the origin of a failure which can lead the system to evolve to any critical failure mode, delaying the instant of a failure by some maintenance actions or simply stopping the system if this is judged necessary.

---

*Corresponding author, E-mail: skadry@gmail.com, Tel: 965-66610985

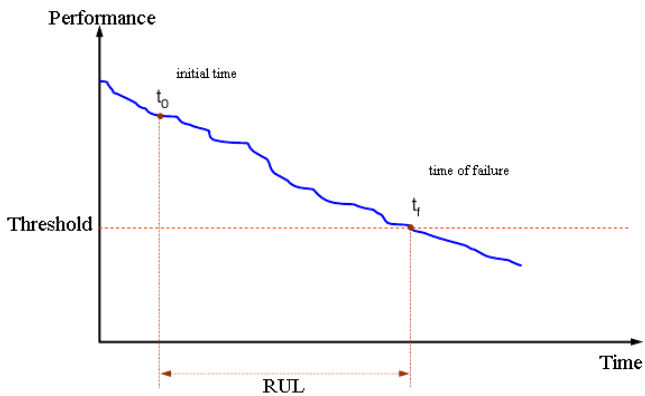

**Figure 2.** Estimation of the value of the RUL

PROGNOSTIC APPROACHES

In the engineering disciplines, fault prognosis has been approached via a variety of techniques ranging from Bayesian estimation and other probabilistic / statistical methods to artificial intelligence tools and methodologies based on notions from the computational intelligence arena. Figure 3 [6, 9] summarizes the range of possible prognosis approaches as a function of the applicability to various systems and their relative implementation cost. Prognosis technologies typically use measured or inferred features, as well as data-driven and/or physics-based models, to predict the condition of the system at some future time. Inherently probabilistic or uncertain in nature, prognosis can be applied to failure modes governed by material condition or by functional loss. Prognosis algorithms can be generic in design but specific in terms of application. Prognosis system developers have implemented various approaches and associated algorithmic libraries for customizing applications that range in fidelity from simple historical/usage models to approaches that use advanced feature analysis or physics-of-failure models.

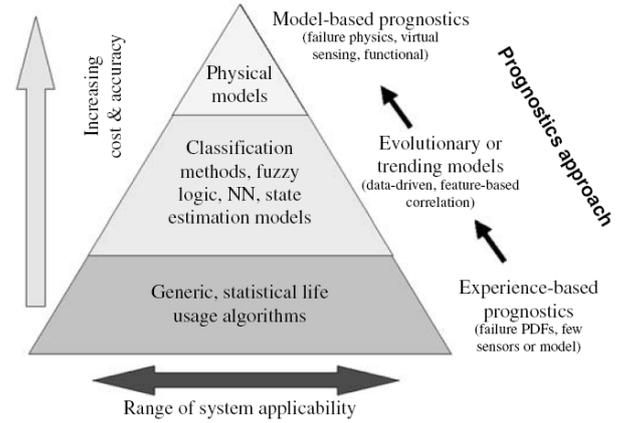

**Figure 3.** Prognosis technical approaches.

Table 1 [9] provides an overview of the recommended models and information necessary for implementing specific approaches. Of course, the resolution of this table only illustrates three levels of algorithms, from the simplest experienced-based (reliability) methods to the most advanced physics-of-failure approaches that are calibrated by sensor data.

**Table 1.** Prognostic accuracy

| Prognostic accuracy | Experienced-Based | Evolutionary | Physics-Based |
|---|---|---|---|
| Engineering model | Not required | Beneficial | Required |
| Failure history | Required | Not required | Beneficial |
| Past operating conditions | Beneficial | Not required | Required |
| Current conditions | Beneficial | Required | Required |
| Identified fault patterns | Not required | Required | Required |
| Maintenance history | Beneficial | Not required | Beneficial |
| In general | No sensors/no model | Sensors/no model | Sensors and model |

*Experience-based prognostic:* it consists of using probabilistic or stochastic models of the degradation phenomenon, or of the life cycle of the components, by taking into account the data and knowledge accumulated by experience during the whole exploitation period of the industrial system. The probabilistic model can be a simple probability function or a modeling in the form of stochastic process. In this framework, the most used probability functions are: Weibull law, exponential law when the failure rate is supposed to be constant, and normal, log-normal and Poisson laws. The parameters of each law are estimated from the data gathered during the whole exploitation period of time (experience feedback, maintenance data, etc.). Stochastic process models can be Markovian or semi-Markovian. The advantage of the methods of this approach is that it is not necessary to have complex mathematical models to do prognostic. Moreover, this approach is easy to apply on systems for which significant data are stored in a same standard that facilitates their use. For example, a company which has conserved during a long period of time a production and maintenance database with some minor rules and standards for data storing, can easily get the estimation of the parameters of the probability laws. However, the main drawback of this approach dwells in the amount of data needed to estimate the parameters of the used laws. Indeed, huge and significant amount of exploitation data are needed in order to determine parameters that model faithfully the degradation phenomenon or the life cycle of the concerned system. Consequently, this approach cannot be applied in the case of new systems for which data from experience feedback do not exist. The other kind of problem is that in most of cases, it is necessary to filter and pre-process the data to extract the useful ones, because the stored data are not always directly exploitable (for example, in the same company, two maintenance operators may enter in two different information or appreciations for the same resolved problem).

*Evolutionary or trending models:* the principle of this approach consists of collecting information and data from the system and projecting them in order to predict the future evolution of some parameters, descriptors or features, and thus, predict the possible probable faults. Without being exhaustive, mathematical tools used in this approach are mainly those used by the artificial intelligence community, namely: temporal prediction series, trend analysis techniques, neuronal networks under all their facets, neuro-fuzzy systems, hidden Markov models and dynamic bayesian networks. The advantage of this approach is that, for a well monitored system, it is possible to predict the future evolution of degradation without any need of prior mathematical model of the degradation. However, the results obtained by this approach suffer from precision, and are sometimes considered as local ones (for the case of neural networks and neuro-fuzzy methods). In addition, the monitoring system must be well designed to insure acceptable prognostic results.

*Model-based prognostic:* this consists of studying each component or sub-system in order to establish for each one of them a mathematical model of the degradation phenomenon. The derived model is then used to predict the future evolution of the degradation ([7], [8], [9]). In this case, the prognostic consists of evolving the degradation model till a determined future instant from the actual deterioration state and by considering the future use conditions of the corresponding component. Three main steps are needed in the framework of model-based prognostic. The first step is related to the construction of an analytical dynamic model including the degradation mechanism or phenomenon, and to the determination of failure thresholds. Follows, in the second step, a setup of a monitoring/diagnostic system which allows to evaluate the actual value of the degradation. Finally, a development or a selection of an adequate technique to solve the derived dynamic model (prediction step) is necessary. The main advantage of this approach dwells in the precision of the obtained results, as the predictions are achieved based on a mathematical model of the degradation. However, the derived degradation model is specific to a particular kind of component or material, and thus, cannot be generalized to all the system components. In addition to that, getting a mathematical model of degradation is not an easy task and needs well instrumented test-benches which can be expensive.

**MOTIVATION**

The most important goal for industry now is to predict the remaining lifetime of their products in order to prevent generally sudden and very expensive failure.

Moreover, the classical strategy based on maintenance is no more efficient and there is a need to a new procedure reflecting the instantaneous evolving state of the product during its functioning.

For that reason, the prognostic approach is more realistic as it is based on an online measurement of the device degradation state.

Many developments in this field are realized where some use non-analytic laws of damage such as the abaci of degradation [10]. To be more realistic, a new analytic prognostic model is proposed here, dealing with the law of degradation by fatigue (Paris' law) in addition to the cumulative law of damage (Miner's law).

This prognostic model is crucial in petrochemical industries for the reason of favorable economic and availability consequences on the exploitation cost [11].

Pipelines tubes are considered as a principal component in petrochemical industries, they are manufactured as cylindrical tubes of radius R and thickness e.

The pipelines failure by fatigue is caused by the fluctuation of the gas pressure-depression along the time t ($0 \leq P \leq P_0$) (Figure 4). These pipelines are unfortunately usually designed for ultimate limits states (resistance)

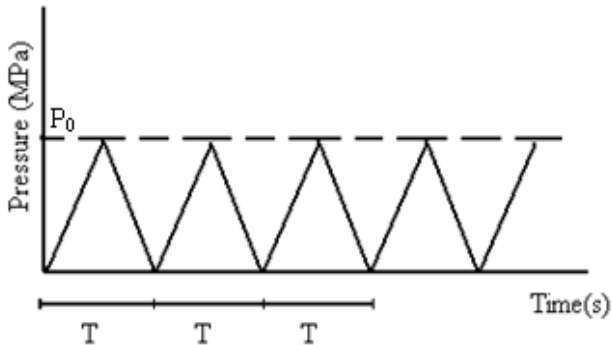

**Figure 4.** Initial pressure variation in pipeline

The fatigue of material under repeated stresses is represented by the Wöhler curve [15] (Figure 5).

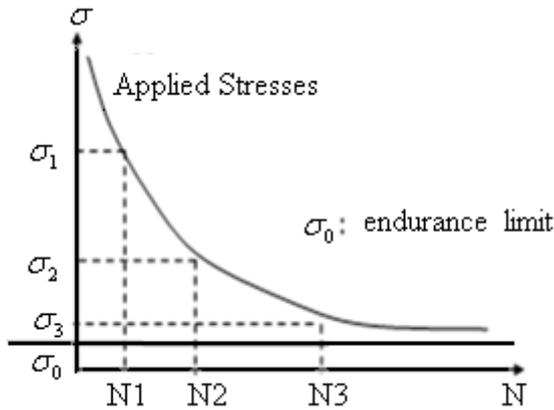

**Figure 5.** Wöhler curve of fatigue

## PARIS LAW

The Paris' law [12, 19] allows determining the propagation speed of the cracks *da/dN* at the time of their detection:

$$\mathrm{d}a/\mathrm{d}N = C.(\Delta K)^m \qquad (1)$$

Where "*a*" is the crack length, *N* is the number of cycles (where the remaining lifetime RUL is derived directly), *C* and *m* are the Paris constants, and $\Delta K$ is the stress intensity factor.
We can distinguish (Figure 6):
- The long cracks that obey to Paris law [13]
- The short cracks that serve to decrease the speed of propagation
- The short physical cracks that serve to increase the speed of propagation

The law can be written also as [13]:
$$\log(\mathrm{d}a/\mathrm{d}N) = \log C + m\log(\Delta K) \qquad (2)$$

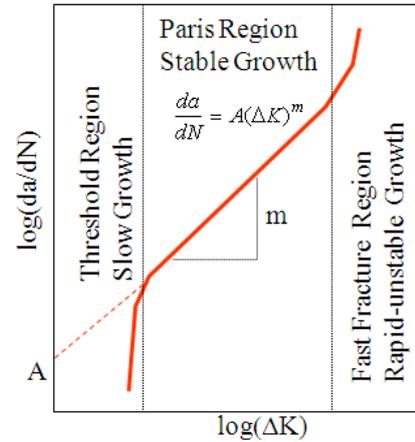

**Figure 6.** The three phases of crack growth, Paris' law

## PIPELINES UNDER PRESSURE

A tube is considered thin when its thickness is of the order of one tenth of its radius: $e \leq R/10$ [14] (Figures 7 & 8)

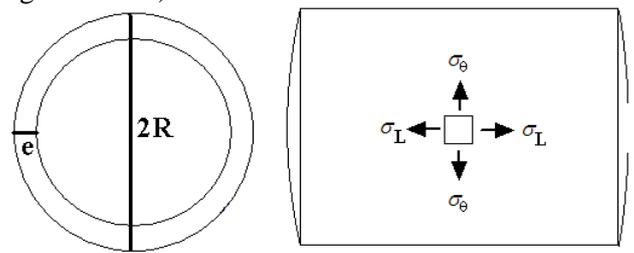

**Figure 7.** Cylindrical pipe dimensions

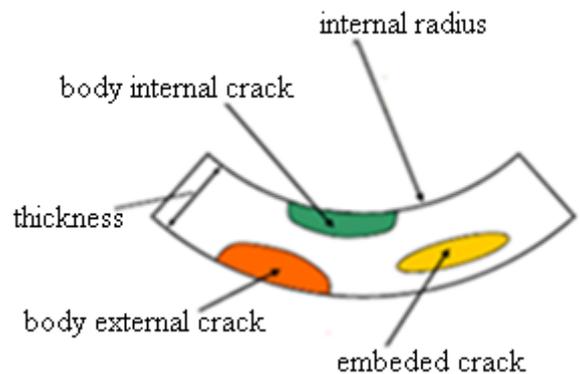

**Figure 8.** Different types of Cracks in pipes

We mention here that we consider only the first mode of crack ($K=K_I$), i.e. the opening mode (the other modes are sliding and tearing mode).

## STATE OF STRESSES IN THE TUBE BODY

The tubes are cylindrical shells of revolution. When thin tubes of radius $R$ and of thickness $e$ are under internal pressure P, the state of stresses is membrane-like without bending loads. The membrane stresses are circumferential (hoop stress) $\sigma_\theta$ and longitudinal stresses (axial stress) $\sigma_L$ (Figure 9).

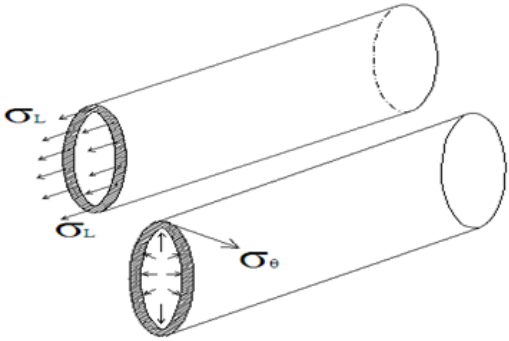

**Figure 9.** Axial and hoop stresses in pipes

These stresses are given by :

$$\begin{cases} \sigma_\theta = (PR)/e \\ \sigma_L = (PR)/(2e) \end{cases} \quad (3)$$

The critical cracks are those which are perpendicular to maximal stresses $\sigma_\theta$, that means longitudinal cracks which are parallel to the axis of the tube. A crack is of depth "$a$" or of length $a$, if we measure in the direction of the tube thickness "$e$". Normally the ratio $a/e$ is within the following range:

$0.1 \leq a/e \leq 0.99$ (Figures 10 & 11).

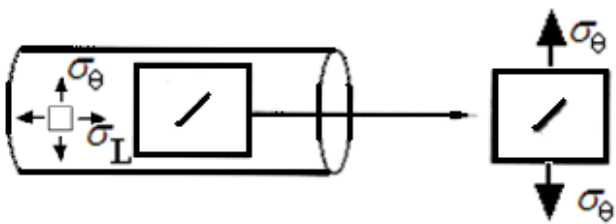

**Figure 10.** Crack length and direction

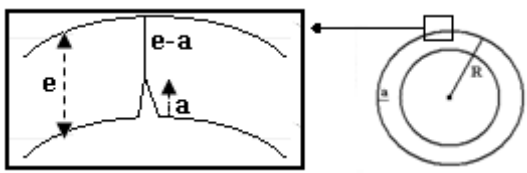

**Figure 11.** Cracked pipe section under internal pressure.

# FAILURE MODE BY SUDDEN CRACK PROPAGATION

The stress intensity factor $K_I$ represents the effect of stress concentration in the presence of a flat crack [15]. In figure 12 we indicate that the uniform distribution of internal stress $\sigma$ in a non cracked piece becomes non uniform in a cracked piece.

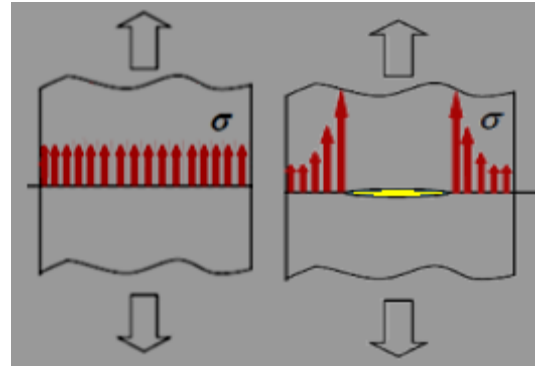

**Figure 12.** Non-uniform distribution of stresses near the crack

The stress intensity factor is given by [16]:

$$K_I = Y(a) \times \sqrt{\pi a}\ \sigma_\theta$$
$$\Rightarrow K_I = 0.6 \times \frac{1 + 2(a/e)}{(1 - a/e)^{\frac{3}{2}}} \times \sqrt{\pi a} \times P.(R/e) \leq K_{IC}$$

With $Y(a)$ is the geometric factor function of the pipeline geometric parameters ($a$, $e$).

$K_{IC}$ : is the tenacity of material (critical stress intensity factor) and is given by:

$$K_{IC} = \sqrt{\frac{J_{IC} \cdot E}{1 - (\nu)^2}}$$

Where:
$J_{IC}$ is the resistant crack force of the material; $E$ is the Young Modulus and $\nu$ is the Poisson ratio. Note that the factor $K_I$ must not exceed the value of $K_{IC}$ [17].

## PROPOSED PROGNOSTIC MODEL

Consider a pipeline of radius $R = 240$ mm and of thickness $e = 8$ mm transporting natural gases, the Paris' law parameters $C$ and $m$ (depend on material, stress ration, temperature, etc. and they are experimentally determined) are taken as being equal to [15]:

$m = 3$ and $C = 5{,}2.10^{-13}$

The length of the crack is denoted by $a$ with an initial value $a_0 = 0.2$ mm

$a_0 \leq a \leq a_N = e/8 \Rightarrow e/a_N = 8$

We have to respect the following ratio:
$0.1 \leq a/e \leq 0.99 \Rightarrow 1.01 \leq e/a \leq 10$

Take a similar form to Paris' law as:
$da/dN = \varepsilon \phi_1(a) \phi_2(p)$

with: $\varepsilon = C$; $\phi_1(a) = \left(Y(a)\sqrt{\pi a}\right)^m$; and

$\phi_2(p) = p^m = (\Delta\sigma)^m$

The initial damage is: $a(0) = a_0$

A recurrent form, which is important to know the historical stages, of crack length gives:

$a_i = a_{i-1} + \varepsilon \phi_1(a_{i-1}) \phi_2(p_i)$  (4)

For the case of natural gas, we consider m=3
$\Rightarrow \phi_2(p_i) = p_i^3 = (\Delta\sigma_{\theta i})^3$

Moreover:
$\eta = \varepsilon/(a_N - a_0)$

We define the damage fraction by:

$d_j = da_j/(a_N - a_0)$

Therefore, we get the cumulated total damage (Figure 13):

$D_i = \sum_{j=1}^{i} d_j = \sum_{j=1}^{i} \frac{da_j}{a_N - a_0} = \frac{\sum_{j=1}^{i} da_j}{a_N - a_0} = \frac{a_i}{a_N - a_0}$

We can easily prove that: $D_N = \sum_{j=1}^{N} d_j = 1$

Where:

$0 \leq n \leq N$, $a_0 \leq a \leq a_N$;

$D_0 \leq D \leq 1 = D_N$; $D_N = \sum_{j=1}^{N} d_j = 1$

$D_0 = \frac{a_0}{a_N - a_0} \Rightarrow a_0 = \frac{D_0 a_N}{1 + D_0}$

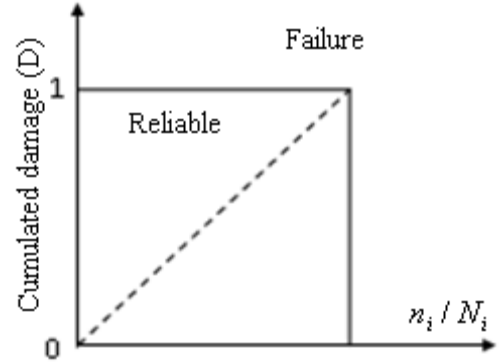

**Figure 13.** Miner's law of damage

The corresponding recurrent form of degradation to equation (4) is given by:

$D_i = D_{i-1} + \eta \phi_1(D_{i-1}) \phi_2(p_i)$

and therefore we can draw the degradation trajectory for the total cycle number of loads.

## SIMULATION OF THREE LEVELS OF INTERNAL PRESSURE

We study by simulation three levels of maximal pressures in pipelines which are: 3 MPa, 5 MPa, and 8 MPa that are repeated within a specific interval of time T = 8 hours. At each level, we deduce the degradation trajectory $D$ in terms of time or in terms of the number of cycles $N$.

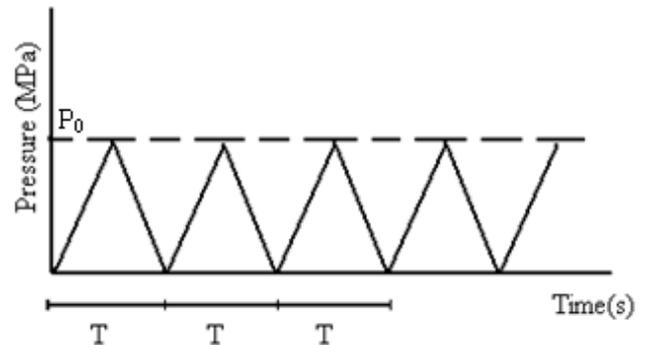

**Figure 14.** Triangular simulation of internal pressure

**Table 2.** Statistical characteristic of each pressure mode

| Pressure Mode | $\bar{p}_i$ **(MPa)** | $\delta p_i$ % | **Law** |
|---|---|---|---|
| High (Mode 1) | 8 | 10% | Triangular |
| Middle (Mode 2) | 5 | 10% | Triangular |
| Low (Mode 3) | 3 | 10% | Triangular |

The failure by fatigue is obtained for a certain critical number of pressure-depression cycles (Figure 14) or for a certain time period. Therefore, the lifetime of the pipeline for each level of maximal pressure is deduced at $D=1$.

In the table 2, we indicate the mean values $\bar{p}_i$ as their maximal values, the coefficient of variation $\delta p_i$, and the law for the three studied pressure levels. These three modes were chosen to represent the middle and the extreme cases of pressure (Maximum, Middle, and Minimum). Their values reflect the real load of petrochemical tubes [18]

## RESULTS OF SIMULATION

The 1000 times Monte Carlo simulations for the pipeline system and under the 3 modes of internal pressure (high, middle and low) gives the degradation trajectories which are represented in the following 3 figures (15, 16 & 17):

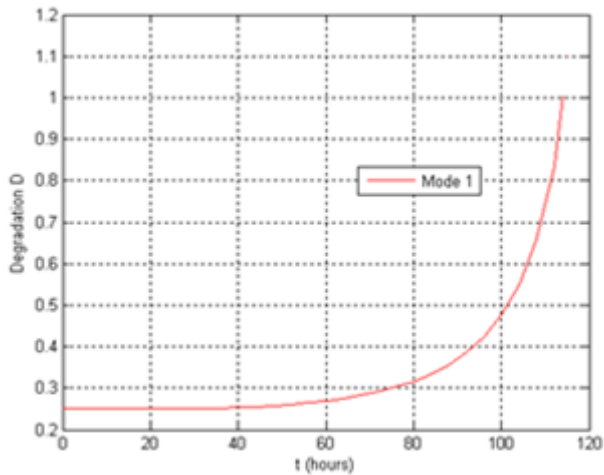

**Figure 15.** Degradation evolution for mode 1

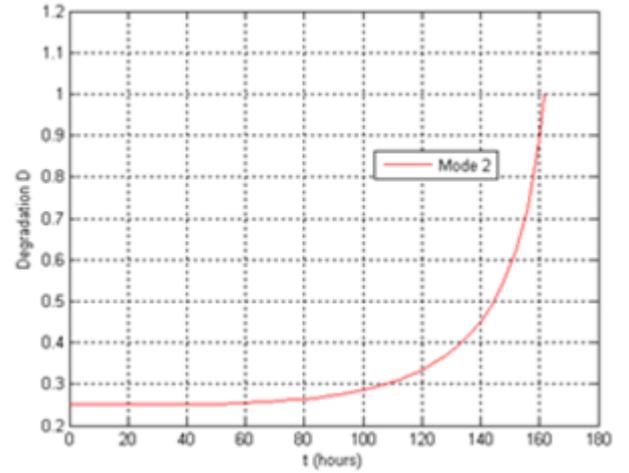

**Figure 16.** Degradation evolution of mode 2

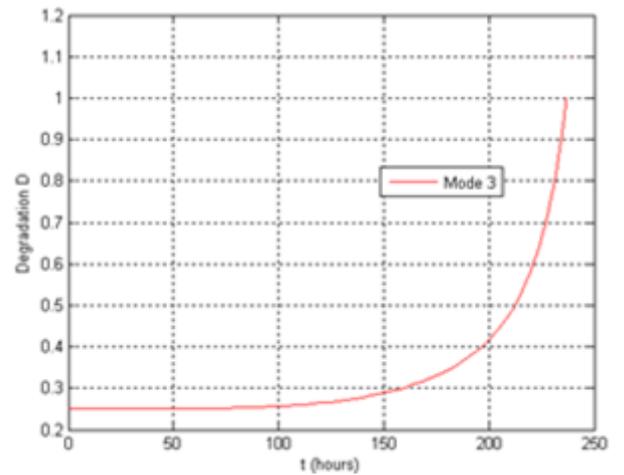

**Figure 17.** Degradation evolution for mode 3

We deduce that the pipeline lifetime is nearly 115 hours for mode 1 (high pressure), nearly 160 hours for mode 2 (middle pressure), and nearly 240 hours for mode 3 (low pressure). From these curves, we can see that our prognostic model, using analytic laws, gives the remaining lifetime or pipelines at any instant. (Figure 18).

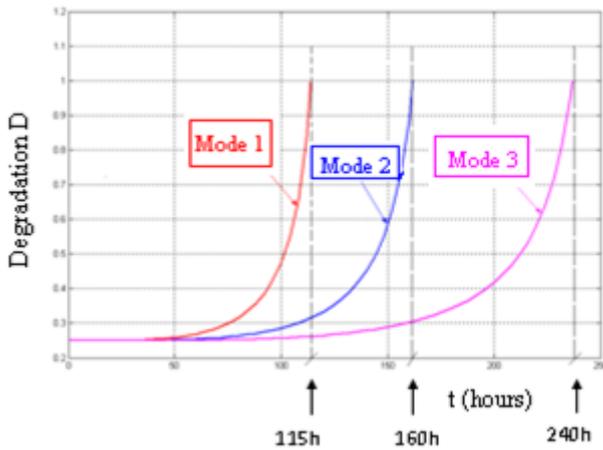

**Figure 18.** Degradation evolution for three modes

## CONCLUSION

A prognostic model is introduced in this paper that permits to predict the degradation trajectory of a pipeline system; it is based on analytical laws such as Paris' law and Miner's law.

This new model is applied to a petrochemical pipeline system and we deduce its degradation trajectory that allows us to determine the remaining useful lifetime which is the main goal of all prognostic studies.

We can notice from the obtained results that the increase of the pipeline lifetime relative to the mode 3 is as follows: mode (1)/mode (3) $\approx 100\%$ and mode (2)/mode (3) $\approx 50\%$.

As a prospective and future work, we intend to more develop our prognostic methodology and apply it to other dynamic systems and this by taking into consideration the probabilistic aspect of basic parameters.